\documentstyle[12pt]{article}
\newfont{\sslarge}{cmss12 scaled \magstep1}
\newfont{\qlarge}{cmbx10 scaled \magstep3}
\newfont{\smr}{cmr12 scaled\magstep1}
\newfont{\bib}{cmb10 scaled\magstep1 }
\newfont{\smaf}{cmr6}
\newcommand{\be}{\begin{equation}}
\newcommand{\bea}{\begin{eqnarray}}
\newcommand{\ee}{\end{equation}}
\newcommand{\ea}{\end{eqnarray}}

\begin{document}

\setlength{\baselineskip}{20pt}

\setcounter{page}{1}
\setcounter{figure}{0}
\setcounter{equation}{0}
\setcounter{section}{0}
\setcounter{subsection}{0}

\centerline{\bf The $\Delta I$=4 bifurcation in
              superdeformed bands }

\vspace{5mm}
\centerline{   P. Magierski$^{a}$, K. Burzy\'nski$^{b}$, }
\centerline{   J. Dobaczewski$^{b}$ and W. Nazarewicz$^{a,b,c}$ }
\vspace{3mm}
\centerline{\em $^{a}$Institute of Physics, Warsaw University of Technology,}
\centerline{\em ul. Koszykowa 75, PL--00-662 Warsaw, Poland}
\vspace{1mm}
\centerline{\em $^{b}$Institute of Theoretical Physics, Warsaw University,}
\centerline{\em ul. Ho\.za 69, PL--00-681 Warsaw, Poland}
\vspace{1mm}
\centerline{\em $^{c}$Joint Institute for Heavy Ion Research and Physics
Division}
\centerline{\em Oak Ridge National Laboratory, Oak Ridge, Tennessee 37831, USA}
\centerline{\em and Department of Physics and Astronomy, University of
Tennessee}
\centerline{\em Knoxville, Tennessee 37996, USA}
\vspace{5mm}

\begin{abstract}
The origin of the $\Delta I$=4 staggering effect which has recently been
observed in several superdeformed bands is discussed.
The Hamamoto and Mottelson model, which is based on a phenomenological
parametrization of the hamiltonian as a quartic function of angular
momentum, is analyzed. A stability of the description with respect
to the $C_4$-symmetry breaking terms is studied.

\end{abstract}


\bigskip

Recently, a puzzling
 new phenomenon
has been observed
in few superdeformed rotational sequences. Namely,
dynamical moments of inertia of several bands
exhibit small regular variations (staggering) as functions of
angular momentum. The effect consists in systematic energy
displacements of rotational states, which are alternately pushed up and
down along the rotational band.
Thus the energy levels are separated into two
sequences with the spin values $I$, $I$+4, $I$+8, $\ldots$ and
$I$+2, $I$+6, $I$+10, $\ldots$, respectively.
One should mention that the effect is of the order of 50 eV,
and since the excitation energies of superdeformed bands
are of the order of
20 MeV, the observed shifts correspond to a $10^{-6}$ perturbation of
the energy levels.

This phenomenon was first observed in the
superdeformed yrast band of $^{149}$Gd \cite{Fli}. To date the staggering
has also been found in three bands of
$^{194}$Hg~\cite{Ced} and in one band of $^{153}$Dy~\cite{Asz}.
The characteristic
feature is that the bifurcation into two
$\Delta I$=4 families  appears at a certain (high) rotational
frequency, the amplitude of the oscillations increases with
spin, and, sometimes, the
change of phase in the oscillatory pattern is observed.

The bifurcation of the rotational band into the two families of states
can be associated with the existence of a higher-order symmetry
in superdeformed states. The symmetry results in the appearance of
a new quantum number distinguishing the two branches.
Since the period of the oscillations is $\triangle I$=4, a
possible explanation can be based on the four-fold rotational
symmetry~\cite{Fli}. The possible origin of this symmetry could be
the coupling between the rotational motion and the hexadecapole vibration.
We expect that for high angular momenta the hexadecapole phonon
becomes aligned along the rotation axis
and gives rise to a small hexadecapole perturbation of the
average field.
The above mechanism was suggested by Peker {\it et al.}
\cite{Pek} to explain a similar effect observed at low spins in
$^{236}$U, $^{238}$U, and $^{218}$Ra nuclei.

In order to study a possibility
that the hexadecapole correlations are
responsible for the staggering in the moment of inertia,
we analyzed the exact
solutions of the single-$j$ shell system with identical particles
interacting via quadrupole-quadrupole and  hexadecapole-hexadecapole
interactions~\cite{Bur}.
The Hamiltonian for this simple model can be written in the following form:
\be
   \hat H = - \chi_2 \hat Q_2 \cdot \hat Q_2
            - \chi_4 \hat Q_4 \cdot \hat Q_4 \;,
\ee
where $\hat Q_\lambda \cdot \hat Q_\lambda$=$\displaystyle{\sum_{\mu}}
(-)^{\mu}\hat Q_{\lambda \mu}\hat Q_{\lambda -\mu}$ and
$\hat Q_{\lambda\mu}$
for $\lambda$=2 and 4
denotes the quadrupole and hexadecapole operator, respectively.
Such a model has two
properties which are of great importance in this case: it fully takes into
account quantum dynamics and yields states with good angular momentum.

We analyzed collective quadrupole bands for
$j$=15/2 and $j$=21/2 and different particle numbers.
We found that the staggering effect (see Fig.~1) appears in yrast bands when
the hexadecapole coupling constant
$\chi_4$ is larger than a certain critical value $\chi_4^{{\rm crit}}$.
This shows that the hexadecapole correlations may indeed be
responsible for the
$\triangle I$=4 staggering.

On a phenomenological level, effects of the hexadecapole deformation
can be simulated by using an intrinsic Hamiltonian which is a quartic function
of angular momentum. Such an operator can in principle
have the same spatial four-fold
symmetry as the microscopic mean field with hexadecapole distorsions.
Recently such an approach
has been proposed by Hamamoto and Mottelson~\cite{HM}.
They considered a rotational Hamiltonian with an explicit four-fold
symmetry around the $z$-axis (perpendicular to the rotational axis)
which has the form:
\be\label{HH}
\hat{H}=A\hat{I}_{z}^{2} + B_{1}(\hat{I}_{x}^{2}-\hat{I}_{y}^{2})^{2}
                     + B_{2}(\hat{I}_{x}^{2}+\hat{I}_{y}^{2})^{2} \;.
\ee
The four lowest states  form then a quartet of
states which can be labeled by the quantum number associated with
the $C_4$ symmetry.
The authors found that
for a certain ratio of parameters $A$/$B_{1}$ and for $B_{2}$=0
the staggering appears as a result of
the tunneling between
the four degenerate different minima in the total energy surface
of Hamiltonian (\ref{HH}).

In the present study we address the
question whether the staggering
occurs also when the $C_4$ term in the Hamiltonian is a small
perturbation on the top of the  large $C_2$-invariant  term
(i.e., term which posesses a two-fold symmetry).
We consider a modified Hamiltonian of the form
\be\label{e3}
\hat{H}=A\hat{I}_{z}^{2} + D_{1}I(I+1)(\hat{I}_{x}^{2}-\hat{I}_{y}^{2})
                         + B_{1}(\hat{I}_{x}^{2}-\hat{I}_{y}^{2})^{2} \;.
\ee
We neglect terms proportional to
$(\hat{I}_{x}^{2}$$+$$\hat{I}_{y}^{2})$ and
$(\hat{I}_{x}^{2}$$+$$\hat{I}_{y}^{2})^{2}$
as they lead only to a renormalization of the coefficient $A$.
The additional term in (\ref{e3}) has the explicit
two-fold symmetry and henceforth  breaks the $C_{4}$ symmetry
when $D_{1}$$\neq $0.
We put parameters $A$ and $B_{1}$ to be equal $90$ and $1$,
respectively. For these values Hamamoto and Mottelson
\cite{HM}  found
the staggering pattern present.
Having the parameters $A$ and $B_{1}$ fixed we changed the
coefficient $D_{1}$ (see Fig.~2).
One can see that even for relatively high values of $D_{1}$ the
staggering effect is still
present.
 Although the $C_{4}$
symmetry is broken in this case, and the corresponding quantum
number has only an approximate meaning, one is still able to
use this label to
distinguish states of the lowest quartet.
Finally, for the chosen set of parameters $A$ and $B_1$ the staggering
vanishes at $D_{1}$$\approx$$1.4$. In this limit, the system
rotates around the $y$-axis (see Fig.~3). This is a consequence of
the appearance of the two well-defined minima in the total energy surface
(see Fig.~4).
The quartet of the lowest states separates then into
two families of degenerate pairs labeled by the quantum
number of $C_{2}$.

The above results suggest that even a small $C_{4}$-type perturbation
can give rise to a staggering effect.
This may suggest that one can search for a microscopic
justification of the four-fold symmetry
with respect to the rotational axis, where the $C_{2}$ term
dominates. One should mention that according to recent
cranked-Nilsson calculations~\cite{Ing} there is no evidence
for a strong
four-fold symmetry breaking with respect to the axial symmetry
axis.

\begin{center}
{\bf ACKNOWLEDGEMENTS}
\end{center}

This research was supported in part by the Polish State Committee for
Scientific Research under Contracts Nos. 20450 91 01 and 2 P302 056 06
and by the computational grant from the Interdisciplinary Centre for
Mathematical and Computational Modelling (ICM) of Warsaw University.
Oak Ridge National
Laboratory is managed for the U.S. Department of Energy by Martin
 Marietta Energy Systems, Inc. under Contract No.
DE-AC05--84OR21400.
The Joint Institute for Heavy Ion
 Research has as member institutions the University of Tennessee,
Vanderbilt University, and the Oak Ridge National Laboratory; it
is supported by the members and by the Department of Energy
through Contract No. DE-FG05-87ER40361 with the University
of Tennessee.  Theoretical nuclear physics research
at the University of Tennessee
 is supported by the U.S. Department of
Energy through Contract No. DE-FG05-93ER40770.

\newpage
Fig. 1 \\
Staggering of transition energies in the yrast band obtained
for 8 particles moving in the single-$j$ shell for $j$=15/2.
Open circles denote results for
pure quadrupole interaction, $\chi_4$=0. Full circles represent results
obtained for the system with perturbation of the hexadecapole type for
$\chi_{4}/\chi_{2}$=0.4. The formula used to derive the staggering
amplitude is given in Ref.~\protect\cite{Fli}.

\bigskip
Fig. 2 \\
   Staggering amplitudes calculated for the band obtained
            by using
            Hamiltonian (\protect\ref{e3}) with four different
            values of the coefficient $D_1$.

\bigskip
Fig. 3 \\
  Average values of components of the angular momentum,
  $\langle\hat{I}_{x}^{2}\rangle/I(I$$+$$1)$,
  $\langle\hat{I}_{y}^{2}\rangle/I(I$$+$$1)$, and
  $\langle\hat{I}_{z}^{2}\rangle/I(I$$+$$1)$,
  as functions of $D_{1}$, calculated at $I = 16$ for the same state as in
   Fig.~2.

\bigskip
Fig. 4 \\
  Total energy surfaces for Hamiltonian~(\protect\ref{e3})
  and for $D_1 = 0$ (a) and $D_1 = 1.47$ (b).
  Spherical angles $\Phi$ and $\Theta$
  describe the orientation of the angular momentum treated as
  a classical variable.
  Minima in the plots are denoted by dots.
  Orientations parallel to the $x$, $y$, and $z$ axes
  correspond to $\Phi = 0^\circ$, $\Theta = 90^\circ$, to
  $\Phi = 90^\circ$, $\Theta = 90^\circ$, and
  to $\Theta = 0^\circ$, respectively.
  Calculations were performed for $I = 16$.

\end{document}